\documentclass[prb,floatfix,aps,twocolumn,showpacs,superscriptaddress]{revtex4-1}
\usepackage[dvips]{graphics}
\usepackage{graphicx}
\usepackage[utf8]{inputenc}
\usepackage{amssymb}
\usepackage{epstopdf}
\usepackage{bm}
\usepackage{dcolumn}
\usepackage{epsfig}
\usepackage{latexsym}
\usepackage{amsmath}
\usepackage{color}
\usepackage{array}
\usepackage{enumerate,dsfont}

\def\XXint#1#2#3{{\setbox0=\hbox{$#1{#2#3}{\int}$ }
\vcenter{\hbox{$#2#3$ }}\kern-.5\wd0}}

\newcommand{\sh}{\mathrm{sh}}
\newcommand{\ch}{\mathrm{ch}}

\begin{document}
\title{High-efficiency resonant amplification of weak magnetic fields for single
spin magnetometry}
\title{High-efficiency resonant amplification of weak magnetic fields for single
spin magnetometry at room temperature}
\author{Luka Trifunovic}
\affiliation{Department of Physics, University of Basel, Klingelbergstrasse 82,
CH-4056 Basel, Switzerland}
\author{Fabio L. Pedrocchi}
\affiliation{JARA Institute for Quantum Information, RWTH Aachen University,
D-52056 Aachen, Germany}
\affiliation{Department of Physics, University of Basel, Klingelbergstrasse 82,
CH-4056 Basel, Switzerland}
\author{Silas Hoffman}
\affiliation{Department of Physics, University of Basel, Klingelbergstrasse 82,
CH-4056 Basel, Switzerland}
\author{Patrick Maletinsky}
\affiliation{Department of Physics, University of Basel, Klingelbergstrasse 82,
CH-4056 Basel, Switzerland}
\author{Amir Yacoby}
\affiliation{Department of Physics, Harvard University, Cambridge MA, 02138,
USA}
\affiliation{Condensed Matter Chair, Department of Physics and Astronomy,
University of Waterloo, Canada}
\author{Daniel Loss}
\affiliation{Department of Physics, University of Basel, Klingelbergstrasse 82,
CH-4056 Basel, Switzerland}
\date{\today}
\begin{abstract}
We demonstrate theoretically that by placing a ferromagnetic particle between a
nitrogen-vacancy (NV) magnetometer and a target spin, the magnetometer
sensitivity is increased dramatically. Specifically, using materials and techniques already
experimentally available, we find that by taking advantage of the ferromagnetic
resonance the minimum magnetic moment that can be measured is smaller by four
orders of magnitude in comparison to current state-of-the-art magnetometers. As
such, our proposed setup is sensitive enough to detect a single nuclear spin at
a distance of $30$~nm from the surface within less than one second of data
acquisition at room temperature. Our proposal opens the door for nanoscale NMR
on biological material under ambient conditions.
\end{abstract}

\maketitle
\section{Introduction}
Magnetic resonance techniques not only provide powerful imaging tools that have
revolutionized medicine, but they have a wide spectrum of applications in other
fields of science like biology, chemistry, neuroscience, and
physics.~\cite{Ernst, Hemmer01022013} In order to resolve structures on the
nanometer scale and thus image individual molecules, however, one needs to go
beyond conventional magnetometric techniques. In particular, standard nuclear
magnetic resonance (NMR) and magnetic resonance imaging (MRI) experiments detect
magnetic fields through the current induced inside a coil according to Faraday's
law; unfortunately induction-based detection is not sensitive enough to allow
resolution in the sub-micrometer regime.~\cite{slides_noninductive_1991} 

Over the last years, a lot of experimental effort has been put into improving
magnetic detection schemes. At present, Hall-sensors and SQUID sensors are
among the most sensitive magnetic field
detectors.~\cite{ramsden_hall-effect_2006,Huber_Gradiometric_2008} Furthermore,
a great deal of success has been achieved with magnetic resonance force
microscopy, where the force between a magnetic tip and the magnetic moment
under investigation is exploited to detect single electron-spins, achieving a
resolution of a few cubic
nanometers.~\cite{Degen_Nanoscale_2009,PoggioNanotechnology,PoggioNatPhys2013}
On the other hand, the very low temperatures that are required in such schemes
represent a considerable drawback to imaging systems in many biological
environments.

NV-center spins also provide very good candidates for magnetometry, boosting
sensitivities up to a few  $\mathrm{nT}/\sqrt{\mathrm{Hz}}$ at room
temperature~\cite{Mamin_Nanoscale_2013,grinolds_subnanometre_2014,Staudacher01022013,0034-4885-77-5-056503,WaldherrG_High_2012,Bassett14082014}
and sub-nanometer spatial resolution, permitting three-dimensional imaging of
nanostructures.~\cite{grinolds_subnanometre_2014} These results are realizable
due to the amazingly long decoherence times of NV-centers at room
temperature and the ability to noninvasively engineer an NV-magnetometer very
close to the magnetic sample.  Although impressive, current state-of-the-art
technology~\cite{MaletinskyP_A_2012} is unable to detect a single nuclear spin;
achieving such sensitivity would revolutionize magnetic imaging in chemical and
biological systems by facilitating atomic resolution of molecules.

In this work, we propose an experimental realization of NV-magnetometers which
could increase present NV-center sensitivities by four orders magnitude at room
temperature; this unprecedented amplification of sensitivity forecasts
magnetometers capable of detecting individual nuclear spins. This can be
achieved by introducing a ferromagnetic particle between the spin that needs to
be detected, which henceforth we call a qubit,~\footnote{We emphasize that we
denote the target magnetic moment by 'qubit' solely for the purpose of
convenience in nomenclature and that our scheme does not rely on the quantum nature
of the magnetic moment we aim to measure.} and the NV-magnetometer. When
excited on resonance by the driven qubit, the macroscopic ferromagnetic spin
begins to precess which, in turn, amplifies the magnetic field felt by the
NV-center. By resonantly addressing the qubit and using a ferromagnetic
resonator as a lever, our setup, in contrast to existing schemes, is
particularly advantageous because, due to the large amplification of
sensitivity, the nuclear spin need not lie within a few nanometers of the
surface~\cite{loretz_nanoscale_2014} but rather can be detectable at a distance
of $30$~nm, and, while related existing schemes rely on the \emph{quantum}
nature of a mediator spin,~\cite{schaffry_proposed_2011} our proposal is
\emph{fully classical}. With these novelties, our scheme provides chemically
sensitive spin detection.

\section{Setup}
\begin{figure}
  \centering\includegraphics[width=\columnwidth]{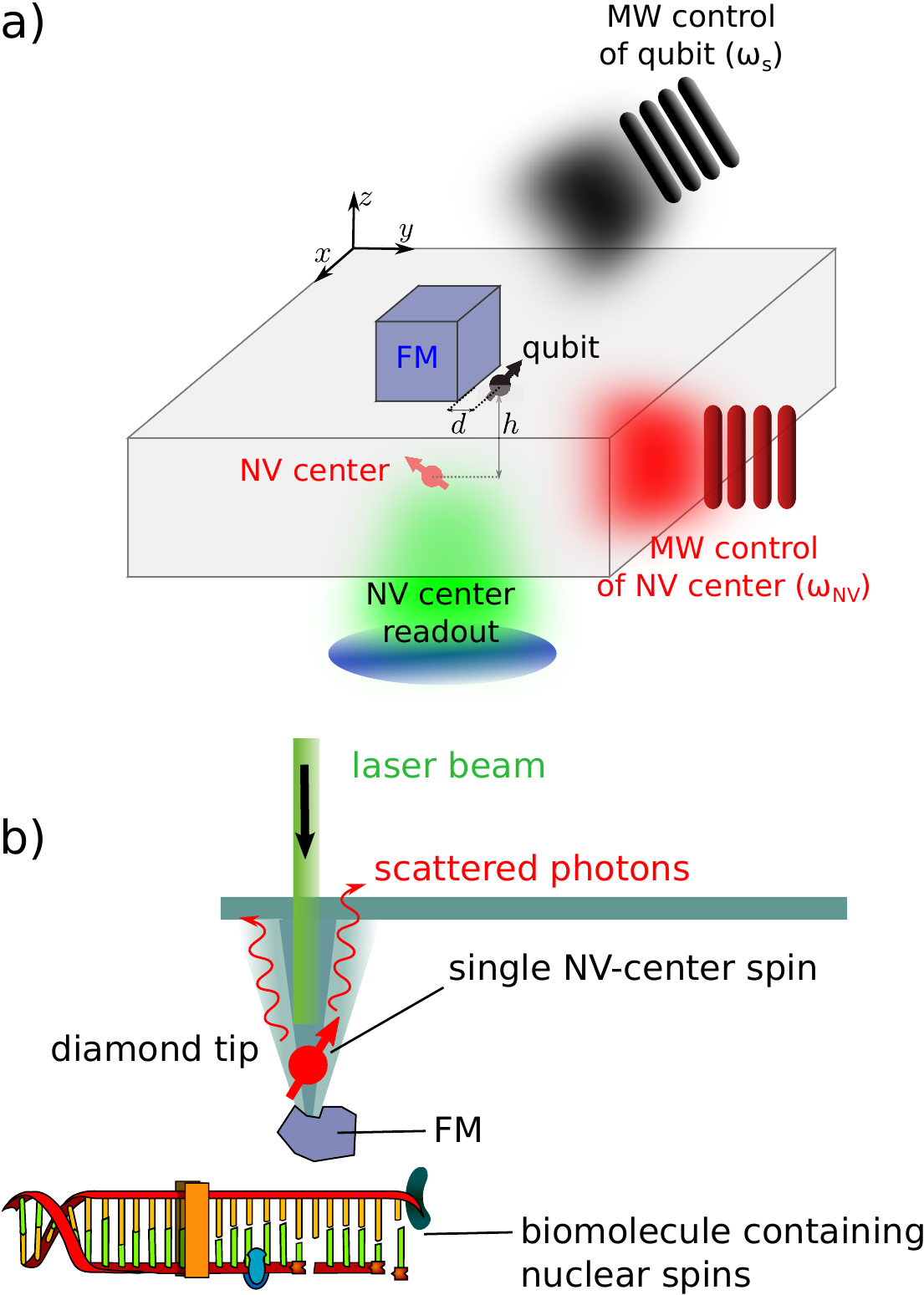}
  \caption{Panel a) shows a detailed illustration of the setup considered.  The
  abbreviation ``FM'' denotes the ferromagnetic particle that is placed on top
  of the diamond surface that contains the NV-center (red) which is used as
  magnetometer. Close to the top surface of the FM lies the qubit
  (black) we want to measure. The setup also includes separate microwave (MW)
  controls of the qubit (black) and NV-center (red) with resonance frequencies $\omega_s$
  and $\omega_\mathrm{NV}$, respectively. The ferromagnetic resonance frequency $\omega_F$
  is assumed to be different from both $\omega_s$ and $\omega_\mathrm{NV}$. The
  NV-center is read out optically with a green laser. A slightly modified version
  of the setup with the NV-center and the FM  on a tip is illustrated in
  panel b); for simplicity we have omitted the two driving fields in this
  panel.}
  \label{fig:setup}
\end{figure}

The standard experimental setup, yielding the most accurate NV-magnetometers
(e.g. Ref.~\onlinecite{grinolds_subnanometre_2014}), consists of an NV-center
near the target qubit and two distinct microwave sources that independently
control the NV-center and qubit so that double electron-electron
(electron-nuclear) resonance, DEER (DENR), can be performed. We extend this
setup by including a macrospin ferromagnetic particle (FM) between the
NV-magnetometer and the qubit we want to measure, see Fig.~\ref{fig:setup}. Due
to the presence of the FM stray field, the qubit energy-splitting, and therefore
the frequency ($\omega_s$) at which the qubit responds resonantly, is strongly
modified; one needs first to characterize the FM stray field in order to be able
to control the qubit by, in our case, applying $\pi$-pulses.\footnote{Instead of
performing the qubit control resonantly, one can make use of `adiabatic
passage'~\cite{Fuchs_A_2011} wherein triangular pulses are applied in lieu of
square pulses. In such a setup, knowledge of the exact value of the qubit Zeeman
splitting, and therefore the FM stray field, is not needed.} Treating the
ferromagnet as a single classical spin, the Hamiltonian of this
system is~\cite{stoner_mechanism_1948,trifunovic_long_2013}
\begin{align}
  \label{eq:H}
  H=&KV(1-m_z^2)+M_FVbm_z-\mu_s\bm n_s(t)\cdot \mathbb{B}_F \bm m,
\end{align}
where $\bm m$ is the normalized magnetization of the FM, $M_F$ the
saturation magnetization of the FM, and $V$ its volume. We assume
uniaxial anisotropy in the FM with the anisotropy constant, $K>0$,
composed of both shape and crystalline anisotropy, with an easy axis along $z$.
An external magnetic field $b$ is applied along the $z$ axis. The magnetic
moment of the qubit is  $\mu_s$ and $\bm n_s(t)$ is its polarization at time
$t$. The $3\times 3$-matrix $\mathbb{B}_F$ is defined as $(\mathbb{B}_F)_{ij}=\bm B_F^j(\bm
r_s)\cdot\bm e_i$, where $\bm B_F^j(\bm r_s)$ is the stray field produced by the
FM at the position of the qubit, $\bm r_s$, when the FM is polarized along the
$j$-axis for $j=x$, $y$, or $z$. The Hamiltonian of the qubit is not explicitly
written as its polarization is completely determined by the applied
time-dependent microwave field and the stray field of the FM. For example, in
equilibrium the ground state of the qubit is polarized along the FM stray field
$\bm n_s=\bm B_F^z/B_F^z$ ($\bm n_s=-\bm B_F^z/B_F^z$) when $m_z=1$ ($m_z=-1$)
and the externally applied magnetic field is small, $\vert b\vert\ll\vert
B_F^z\vert$. Although in the following we take $V$ small enough to approximate
the FM as a monodomain, our analysis and therefore our results are amenable to
including the effects of magnetic texture.

Using two independent microwave sources we apply a train of $\pi$-pulses first
to the qubit and subsequently a Carr-Purcell-Meiboom-Gill (CPMG) pulse
sequence~\cite{cywinski_how_2008,de_Lange01102010} to the NV-center, see
Fig.~\ref{fig:sequence}. As the qubit is pulsed it will drive the FM at the
frequency of the pulse sequence $\pi/\tau$, $\tau$ being the time between the
application of two subsequent $\pi$-pulses. When $\pi/\tau$ is close to the
ferromagnetic resonance (FMR) frequency, $\omega_F$, the response of the FM
becomes large and one obtains a large amplification of the magnetic field felt
by the NV-center. The pulses 
are applied to the qubit only until the FM
reaches steady state precession. We also allow for a possible time offset,
$\xi$, between the pulse sequences applied to the qubit and the NV-center, see
Fig.~\ref{fig:sequence}. Here, $\xi$ may be chosen to compensate for the phase
difference between the  driving of the qubit and the response of the FM, thus maximizing the
amplification. Since the microwave field applied to the qubit is a sequence of
$\pi$-pulses, the polarization is simply $\bm n_s(t)=\bm n_s f_\tau(t)$, where
$f_\tau(t)$ may take the values $\pm1$ according to the pulse sequence. It is
worth noting that even though we excite the FMR with the inhomogeneous dipolar
field of the qubit, only the lowest Kittel mode is excited since for small FM
higher modes are separated by an energy gap that exceeds the perturbation
amplitude. Therefore the macrospin approximation used in Eq.~(\ref{eq:H}) is
justified.

\section{Amplification}
\begin{figure}
  \centering\includegraphics[width=\columnwidth]{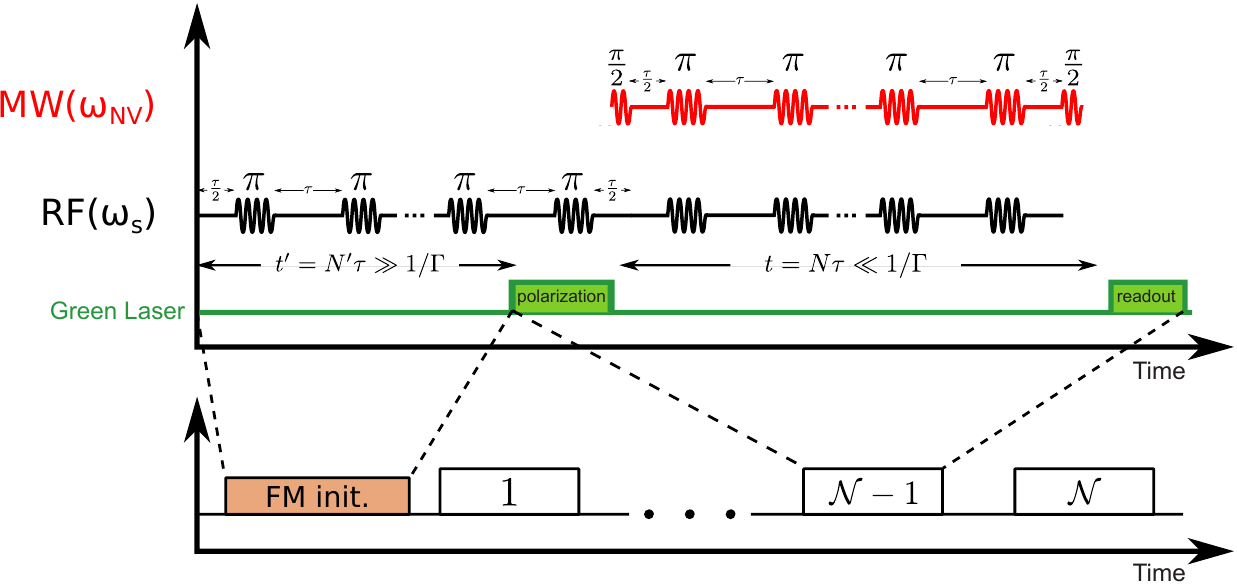}
  \caption{The pulse sequence that we apply to the qubit (black) and to the NV-center
  spin (red). The pulse sequence $f_\tau(t)$ that consists of $N$ ($N$ is even) is
  applied to both spins, with the time offset $\xi$, during the interrogation
  time $t_i=N\tau$. The measurement is repeated $\cal N$ times until the desired
  precision is achieved, as illustrated on the bottom panel. The sequence
  section denoted by ``FM init'' with duration $t^\prime=N^\prime\tau$ is the
  time during which the precession of the FM is being developed. We assume that
  the frequencies $\omega_s$, $\omega_\mathrm{NV}$, and $\omega_F$ are all
  sufficiently different from each other. The green laser is applied to the NV-center for initialization
  (polarization) and read-out. The total measurement time is $t^\prime+{\cal
  N}t_i\approx{\cal N}t_i$.}
  \label{fig:sequence}
\end{figure}

We now consider our particular scenario wherein a FM is introduced at a distance
$d$ from the qubit and $h$ from the NV-center (Fig.~\ref{fig:setup}). In this
case, both the accumulated phase and the dephasing of the NV-center are modified
by the presence of the FM. Because our amplification crucially depends on the
series of pulses applied to the NV-center and qubit, here we detail the pulse
sequence, see Fig.~\ref{fig:sequence}. First we apply, on the qubit only, $N'$
$\pi$-pulses separated by a time interval $\tau$, for a total time of
$t^\prime=N'\tau$---during this time the FM reaches steady state precession.
Next we initialize the NV-center in state $\vert0\rangle$, which takes time
$t_p$. Then, a $\pi/2$ pulse is applied to the NV-center allowing it to
accumulate the phase from the FM tilt stray field. Consequently, a series of $N$
$\pi$-pulses are applied to both the NV-center and qubit for a total
interrogation time $t_i=N\tau$. Finally we apply to the NV-center a
$\pi/2$-pulse which is, in general, along an axis in the plane orthogonal to the
NV-center axis and different from the first $\pi/2$-pulse by an angle $\theta$.
The probability that the NV-center occupies the state $\vert0\rangle$ or
$\vert1\rangle$ after the pulse sequence is now a function of the accumulated
phase $\varphi_\mathrm{NV}(t_i)$
\begin{align}
  p(n\vert\varphi_\mathrm{NV}(t_i))=\frac12\left(1+n\cos(\varphi_\mathrm{NV}(t_i)+\theta)e^{-\langle(\delta\varphi_\mathrm{NV}(t_i))^2\rangle}\right).
  \label{eq:pnphi}
\end{align}
Here, $n=\pm1$ are the two possible outcomes when the state of the NV-center is
measured, $\langle(\delta\varphi_\mathrm{NV}(t_i))^2\rangle$ is the dephasing of
the NV-center, and $\langle\,\cdots\rangle$ is the expectation value in the Gibbs
state. Because the accumulated phase itself depends on the value of the qubit
magnetic moment $\mu_s$, a measurement of the NV-center is a measurement of
$\mu_s$. The variance in the measured value of the NV-center can be reduced by
repeating the measurement $\mathcal{N}$ times (Fig.~\ref{fig:sequence}). Because
typically $t^\prime\ll{\cal N}t_i$ and therefore $t^\prime+{\cal
N}t_i\approx{\cal N}t_i$, the total measurement time is marginally prolonged by
the initial pulse sequence that initialized the tilt of the FM.

Given Eq.~(\ref{eq:pnphi}), one may show quite generally that, in the relevant
experimental limit when $t_i \gg t_p$, the AC sensitivity of the NV-magnetometer
is given by
\begin{equation}\label{eq:Sensitivity}
  S=\frac1{R\sqrt{\eta}}\min_{t_i}\left[\frac{
  e^{\langle(\delta\varphi_\mathrm{NV}(t_i))^2\rangle}\sqrt{t_i}}{\vert
    \partial\varphi_\mathrm{NV}(t_i)/\partial\mu_s\vert}\right]\,,
\end{equation}
which defines the minimum detectable magnetic field for a given total
measurement time. Here, $R$, the measurement contrast, is the relative
difference in detected signal depending on spin-projection of the NV-center
spin, and $\eta$ is the detection efficiency which takes into account that many
measurements have to be performed in order to detect a
photon.~\cite{PhysRevB.80.115202} A detailed derivation of
Eq.~(\ref{eq:Sensitivity}) can be found in Appendix~\ref{sec:Sensitivity_App}.
The sensitivity is small ({\it i.e.}, `good') when the NV-center dephasing is
small while the accumulated phase is large. When the qubit is directly coupled
to the NV-center (unamplified) the dephasing time of the NV-center is given by
$T_2\sim200$~$\mu$s~\cite{ohno_engineering_2012,myers_probing_2014} so that
$\langle(\delta\varphi_\mathrm{NV}(t_i))^2\rangle=(t_i/T_2)^2$.

As we show in Appendix~\ref{sec:cal_App}, given the pulse sequence described
above, when $\Gamma t'\gg 1$ and $\Gamma t_i\ll 1$, where $\Gamma$ is the
linewidth of the ferromagnet, there is a \textit{resonant} response of the FM
while the NV-center picks up \textit{non-resonant} noise. As such, the ratio of
the dephasing to the accumulated phase of the qubit is minimized thereby
optimizing the sensitivity. We henceforth take $\Gamma t'\gg 1\gg\Gamma t_i$ in
the remainder of the text.

The accumulated phase is formally
\begin{equation}
  \varphi_\mathrm{NV}(t_i)=\gamma_\mathrm{NV}\int_0^{t_i} B_\mathrm{NV}(t^{\prime\prime})
  f_\tau(t^{\prime\prime}) dt^{\prime\prime}\,,
  \label{eq:phiNVint}
\end{equation}
where $\gamma_\mathrm{NV}$ is the gyromagnetic ratio of the NV.
$B_\mathrm{NV}\equiv\left\vert\bm B_{F,\mathrm{NV}}^{-}\cdot\bm
n_\mathrm{NV}\right\vert$ where $\bm B_{F,\mathrm{NV}}^\pm=\bm B_F^x(\bm
r_{\mathrm{NV}})\pm i\bm B_F^y(\bm r_{\mathrm{NV}})$ [$\bm B_{F,s}^\pm=\bm
B_F^x(\bm r_{s})\pm i\bm B_F^y(\bm r_{s})$] is a complex combination of the
magnetic stray-field for the FM polarization along the $x$ and $y$ axes at the
position of the NV-center (qubit), $\bm r_{NV}$ ($\bm r_{s}$), and $\bm
n_\mathrm{NV}$ is the NV-center polarization axis. Within the linear response
regime and using the pulse sequence described above and optimally choosing
$\xi$, the expression for the phase accumulated by the NV-center when
$\tau=(2k+1)\pi/\omega_F$,~\footnote{The Fourier transform of the CPMG pulse
sequence has peaks at frequencies $(2k+1)\pi/\tau$} for $k=0,1,\ldots$, is 
\begin{align}
  \varphi_\mathrm{NV}(t_i)&=\frac{4\mu_s\gamma\gamma_\mathrm{NV}
  \vert\bm B_{F,s}^+\cdot\bm n_s\vert\vert\bm B_{F,\mathrm{NV}}^-\cdot \bm
  n_\mathrm{NV}\vert}{\pi^2(2k+1)^2M_FV\Gamma}t_i\,,
  \label{eq:phiNVts}
\end{align}
where $\gamma$ is the gyromagnetic ratio of the FM. $k$ is defined such that the
resonantly driven FM undergoes $2k+1$ half-periods between consecutive
$\pi$-pulses applied to the NV-center. In the optimal case we have $k=0$ so that
$\tau$ is half the period of precession of the ferromagnet. The details of the
derivation of Eq.~\ref{eq:phiNVts} can by found in Appendix~\ref{sec:cal_App}.
It is readily observed from the above equation that
$\varphi_\mathrm{NV}(t_i)\sim1/\Gamma$ which is proportional to the AC magnetic
susceptibility of the FM  on resonance; thus we indeed obtain a resonant
response as anticipated. Even though the phase $\varphi_\mathrm{NV}$ accumulated
due to the FM tilt is large, the angle of the FM tilt is small ($\sim10^{-3}$ if
the qubit is a nuclear spin) because $M_FV\gg\mu_s$. Therefore, we can neglect
the effects of the backaction of the FM tilt on the qubit, because the stray
field modulation induced by the tilt is small compared to the qubit Rabi
amplitude and far detuned from the qubit Larmor precession frequency (i.e.
$\omega_F\neq\omega_s$). Thus, the qubit is polarized along the FM stray field
$\bm n_s=\bm B_F^z/B_F^z$; the scalar product $\bm B_{F,s}^+\cdot\bm n_s$ is
nonzero only if the stray field of the FM tilt has a component along $\bm n_s$
at the position of the qubit. We address the optimal geometry and position of
the qubit relative to the FM in Sec.~\ref{sec:geometry}. 

The relevant dephasing is the maximum of the inherent dephasing of the
NV-center, $(t_i/T_2)^2$, and the dephasing due to the coupling to the
FM,~\cite{cywinski_how_2008}
\begin{equation}
  \beta(t_i,\tau)=\gamma_\mathrm{NV}^2\int_0^{t_i}ds
  S(s)\int_0^{t_i-s}dt^{\prime\prime}
  f_\tau(t^{\prime\prime})f_\tau(t^{\prime\prime}+s).
  \label{eq:dephasin}
\end{equation}
Here $S(s)=\langle B_\mathrm{NV}(s)B_\mathrm{NV}(0)\rangle$ is the
autocorrelation function of the FM noise. Again taking
$\tau=(2k+1)\pi/\omega_F$, we show in Appendix~\ref{sec:D2} that
\begin{align}
  \beta(t_i,\tau)&=\frac{4\gamma\gamma_\mathrm{NV}^2\vert
  \bm B_{F,\mathrm{NV}}^+\cdot \bm
  n_\mathrm{NV}\vert^2k_BT}{\pi^2(2k+1)^2M_FV\omega_F}t_i^2\equiv(t_i/T_2^\prime)^2\,.
  \label{eq:deltaphi}
\end{align}
Because $\beta(t_i,\tau)\sim1/\omega_F\sim S(\omega=0)$, the NV-center indeed
accumulates non-resonant noise. 

After substituting
$\langle(\delta\varphi_\mathrm{NV}(t_i))^2\rangle=\max[\frac{t_i}{T_2},\beta(t_i,\tau)]$
and $\varphi_\mathrm{NV}(t_i)$ from Eq.~(\ref{eq:phiNVts}) and
Eq.~(\ref{eq:deltaphi}) into Eq.~(\ref{eq:Sensitivity}) and performing the
optimization over the interrogation time in Eq.~(\ref{eq:Sensitivity}), we find the
ratio of the unamplified to the amplified sensitivity
\begin{align}
  \label{eq:eta}
  \nu&\equiv\frac{S_U}{S_A}\\
  &= \frac{\sqrt{8}e^\frac14\mu_s\gamma
    \vert\bm B_{F,s}^+\cdot\bm n_s\vert\vert\bm B_{F,\mathrm{NV}}^-\cdot \bm
      n_\mathrm{NV}\vert^\frac12}{\pi^\frac32(2k+1)^\frac32M_FV\Gamma
      B_d\sqrt{\gamma_\mathrm{NV} T_2}}\left(\frac{M_FV\omega_F}{\gamma
    k_BT}\right)^\frac14\nonumber,
\end{align}
where $B_d=\mu_0\mu_s/[4\pi(d^2+h^2)^\frac32]$ is the dipolar field of the qubit
at the position of the NV-center in the unamplified case and $T_2$ is the
NV-center decoherence time when the FM is not present; these quantities define
the unamplified sensitivity. The biggest amplification is obtained when one
half-period of the FM oscillation occurs over the timescale $\tau$, {\it i.e.},
$k=0$. In practice, experimental limitations, such as limitations to the qubit Rabi
frequency, bound $\tau$ and therefore $k$ from below. Thus, in order to
achieve the resonance, one has to use $k\gg0$ (at the expense of sensitivity) or
to tune the FMR frequency as described in the following subsection.

\subsection{Tuning the FMR frequency}\label{sec:tuningFMR}
It has been demonstrated experimentally~\cite{fuchs_gigahertz_2009} that the
electron spin of NV-centers can be coherently driven at GHz frequency. For a
proton spin, however, the same drive would yield Rabi oscillations in the
MHz range. Because typical FMR frequencies are in GHz range, $\omega_{F}$
needs to be reduced in order for the proton Rabi frequency to be on resonance
with the FMR.
\begin{figure}
  \centering\includegraphics[width=\columnwidth]{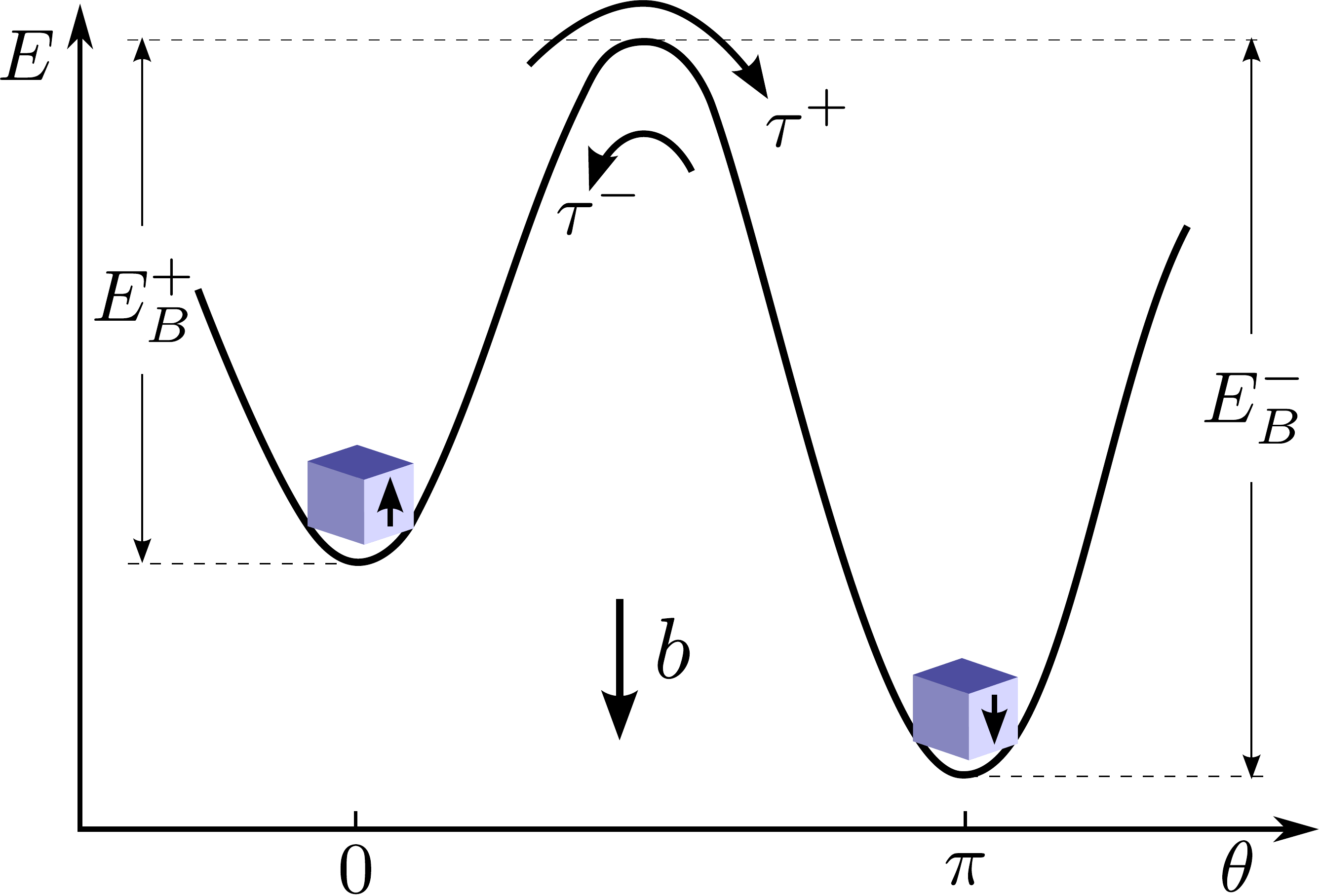}
  \caption{The FM energy when an external field $b/b_a=0.2$ is applied,
  {\it i.e.}, the first two terms from the right-hand side of Eq.~(\ref{eq:H}), as a
  function of $\theta$, where $m_z=\cos\theta$. The metastable state at $\theta=0$ has smaller
  FMR frequency compared to the case with no external field. The tunneling time
  $\tau^+$ from the metastable state has to be longer than the measurement time.
  We note that $E_B^+=M_FV(b_a-b)$.}
  \label{fig:DWpotential}
\end{figure}

One way to decrease $\omega_F$ is to apply an external magnetic field
antiparallel to $\bm m$,~\cite{smith_static_1958} whereby there is a metastable
state when $b<b_a$, with $b_a=K/M_F$ the FM (crystalline and
shape) anisotropy field. In Fig.~\ref{fig:DWpotential}, we plot the energy of the FM
as a function of angle $\theta$ of the magnetization with respect to the easy
axis, according to Eq.~(\ref{eq:H}). It is straightforward to show that the FMR
frequency in the metastable state is $\omega_F^+=\gamma(b_a-b)$. On the other
hand, the ferromagnet will relax to the thermal state on a timescale $\tau^+$
given by the Arrhenius law $\tau^+=\tau_0 e^{E_B^+/k_BT}$, where
$\tau_0\sim1/\omega_F^+$ is the attempt time. We can insure that the FM is
initialized in the metastable state by first measuring the direction of the
magnetization, applying an external magnetic field $\bm b$ antiparallel to $\bm
m$ and checking subsequently that the FM magnetization direction is unchanged,
which can be done under a
nanosecond.~\cite{Julliere1975225,Lee_excitations_2004} In order for the
ferromagnet to remain in the metastable state while the measurement is being
performed, we require $\tau^+\gg1/\Gamma$.  Indeed, the total measurement time
$T$ should be larger than the FMR initialization time $t^\prime\gg1/\Gamma$, and
smaller than Arrhenius' timescale $\tau^+\gtrsim T$, see
Fig.~\ref{fig:sequence}. Thus, if we want to tune $\omega_F^+$ to a certain value
and work at room temperature, the Arrhenius law suggests that the FM 
volume must satisfy
\begin{equation}
  V\gtrsim \frac{\gamma k_BT}{M_F\omega_F^+}\vert\ln\alpha\vert,
  \label{eq:volumineq}
\end{equation}
in order for the metastable state lifetime to be bigger than the measurement
time. Here $\alpha=\Gamma/\omega_F^+$ is the Gilbert damping of the
FM. Substituting Eq.~(\ref{eq:volumineq}) for the minimal
volume into Eq.~(\ref{eq:eta}) we obtain

\begin{align}
  \nu=& \frac{\sqrt{8}e^\frac14\mu_s \vert\bm B_{F,s}^+\cdot\bm n_s\vert\vert\bm B_{F,\mathrm{NV}}^-\cdot \bm
    n_\mathrm{NV}\vert^\frac12}{\pi^\frac32(2k+1)^\frac32\alpha
    k_BT\vert\ln\alpha\vert^\frac34B_d\sqrt{\gamma_\mathrm{NV}T_2}}\,.
  \label{eq:etaV}
\end{align}
Compared to the amplification formula in Eq.~(\ref{eq:eta}), the above equation
is independent of the FMR frequency $\omega_F$ and the FM volume $V$. Thus, irrespective of the
choice of the frequency we work at, the same value for the amplification is
obtained. Furthermore, the only dependence on the volume is incorporated in the
stray fields but, as shown in Sec.~\ref{sec:geometry}, this dependence is weak
in the limit $d,h\ll V^{1/3}\equiv L$. The volume in Eq.~(\ref{eq:volumineq}) is
implicitly bounded from above in order to remain in the regime where the macrospin
approximation is valid. As detailed in Sec.~\ref{sec:est}, FMs with
volumes corresponding to MHz resonance at room temperature are well-described by
a single classical spin.

\subsection{FM  geometry and  demagnetizing
fields}\label{sec:geometry}
\begin{figure}
  \centering\includegraphics[width=\columnwidth]{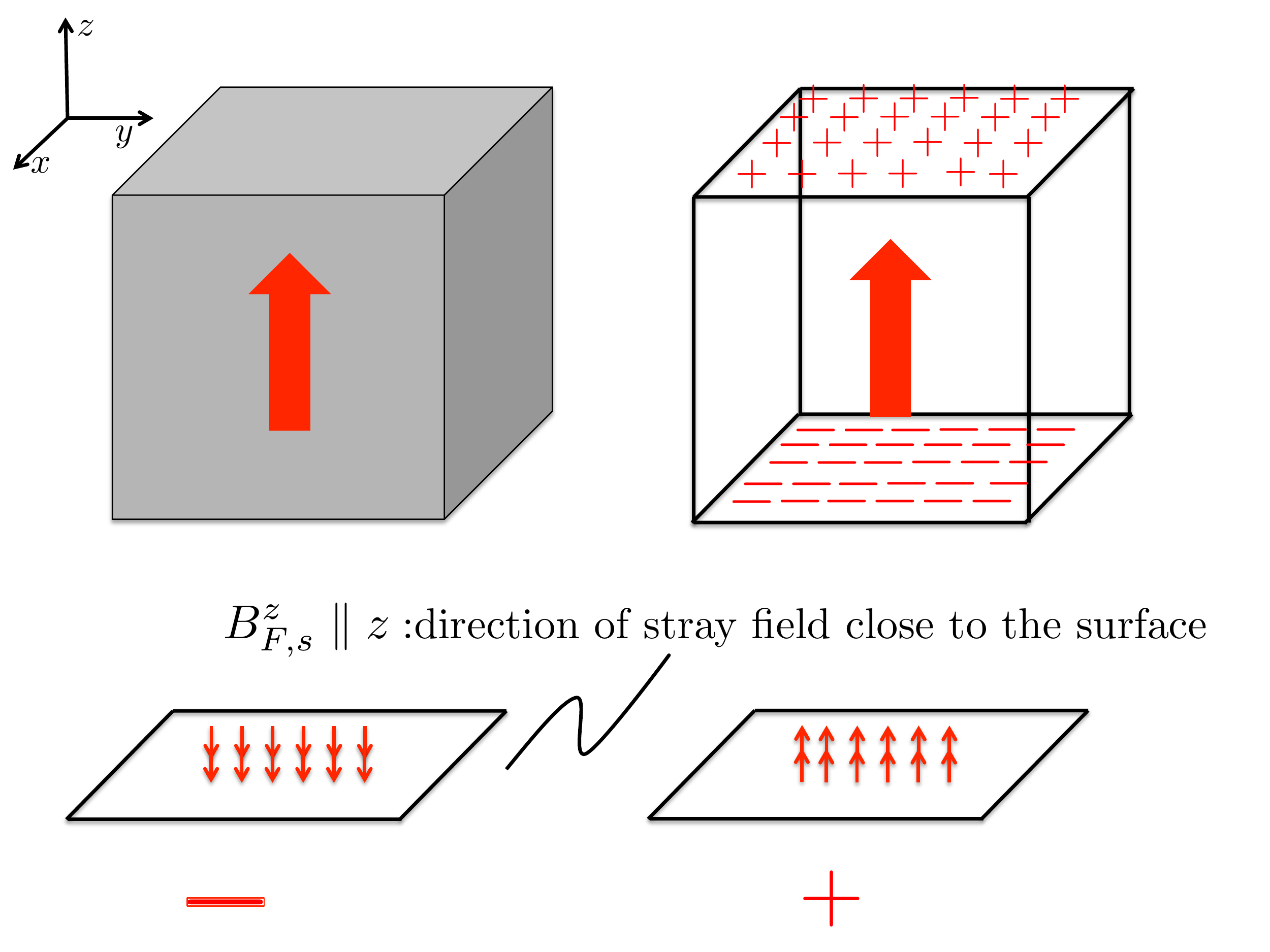}
  \caption{Schematic representation of the ferromagnetic particle polarized
  along the $z$ axis. The stray field produced by a uniformly magnetized cube can be
  calculated by adding the electric field produced from the negatively charged
  bottom plane to the electric field produced by the
  positively charged upper plane. Close to the surfaces and away from the edges,
  the stray field points mostly along $z$. Only close to the edges the
  transverse components become significant.}
  \label{fig:geometry}
\end{figure}

\begin{figure}
  \centering\includegraphics[width=\columnwidth]{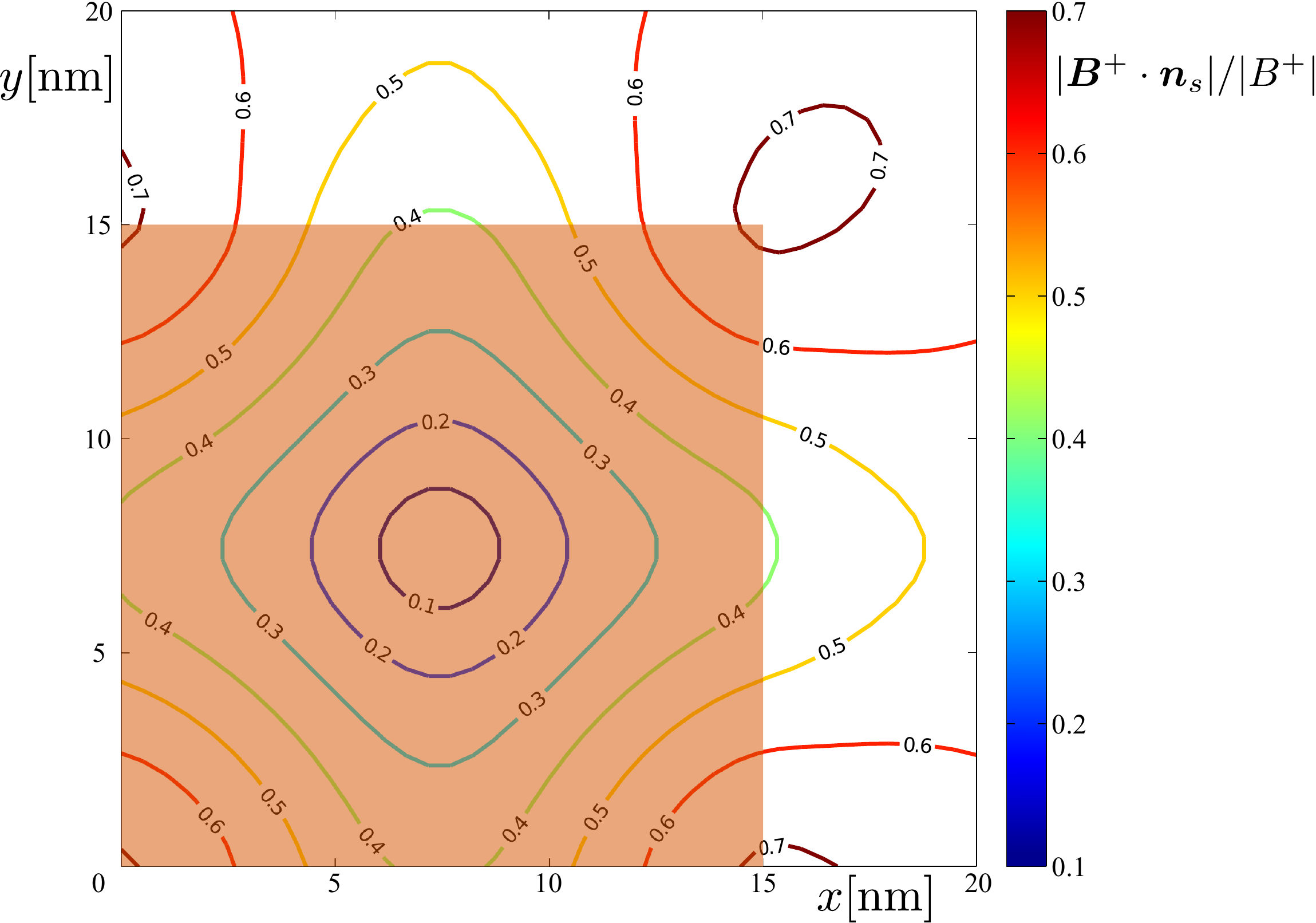}
  \caption{Contour plot of the quantity $\vert\bm B^+\cdot\bm n_s\vert/\vert
  B^+\vert$ in the $xy$-plane that is $2$~nm above the upper face of the cube. We
  assume the FM cube (orange square) has a side length of $L=15$~nm. The
  values of the stray fields are obtained from OOMMF micromagnetic simulations,
  taking into account the demagnetizing field.}
  \label{fig:strayfield}
\end{figure}

In the absence of an external magnetic field, the qubit aligns along the stray
field direction of the FM, while the FM spins are aligned along the easy
axis, according to Eq.~(\ref{eq:H}). Because $M_FV\gg\mu_s$, the FM
tilt induced by the qubit is negligible. Therefore, the qubit will align along
the direction of the stray field produced by the FM. However, for most
geometries of the FM and positions of the qubit, $\bm B^{x,y}_{F,s}\cdot\bm
B^z_{F,s}\sim0$, and therefore the amplification $\nu\sim0$. In the following
discussion, we consider our ferromagnet to be a cube of side $L$, but our conclusions
can be straightforwardly generalized to other geometries. To gain insight into
the direction and strength of the stray field, we use the well-known analogy
between the stray field of a homogeneously magnetized body and an electric field
produced by surface charges,~\cite{samofalov_strong_2013} see
Fig.~\ref{fig:geometry}. Specifically, we may consider the surfaces of the cube
to have charge density $\sim M_F\,\bm m\cdot\bm s$, where $\bm s$ is the vector
normal to the surface of the cube. Therefore, when the position of the qubit is
very close to the center of the FM surface which is perpendicular to the
polarization direction (here assumed along $z$-axis), $\bm B^z_{F,s}$ points
along the $z$-axis, see Fig.~\ref{fig:geometry}.  Similarly, $\bm B^{x}_{F,s}$
and $\bm B^{y}_{F,s}$ are almost aligned with the $x$ and $y$ axes close to the
surface, respectively. Therefore, in these positions, $\bm B^{x,y}_{F,s}\cdot\bm
B^z_{F,s}\sim0$. However, this is not true near the edges of the ferromagnet.
Therefore, in order to obtain a strong amplification, one needs, first, a
ferromagnet with edges and, second, to position the qubit close to the edges.
One may show analytically and numerically (see Fig.~\ref{fig:strayfield}) that
$\vert\bm B^+\cdot\bm n_s\vert/\vert B^+\vert$ close to the edges is about an
order of magnitude bigger than close to the face center and that it has local
maxima close to the cube's corners.

In evaluating $\bm B^{x,y,z}_{F,s}$, we assume that the FM is homogeneously
magnetized as, in cubic geometry, one can find an analytical formula for the
stray field in this case, see Appendix~\ref{app:Stray}. However, it is important
to note that due to demagnetizing fields (arising from dipole-dipole
interactions in the FM), the FM ground-state is not homogeneous but rather
``flowerlike''.~\cite{aharoni_introduction_2001} Specifically, the canting of the
spins close to the edges is more pronounced,~\cite{samofalov_strong_2013} which
modifies the FM stray field close to the edges. To account for the effects of
the demagnetizing fields, we perform micromagnetic simulations in
OOMMF.~\footnote{The code is available at http://math.nist.gov/oommf} In
Fig.~\ref{fig:strayfield} we plot $\vert\bm B^+\cdot\bm n_s\vert/\vert B^+\vert$
in the $xy$-plane that is $2$~nm above the upper face of the cube. We find that
the inclusion of demagnetizing fields changes our value of $\bm B^{x,y,z}_{F,s}$
by only $\sim1\%$ as compared to the uniformly magnetized cube. Therefore, we
expect the analytical expression for the stray field to be valid for our choice
of parameters.

Because the amplification depends on $h$ and $d$ only through the stray fields,
here we detail this dependence and show that our scheme is robust against small
variations of $h$ and $d$. The stray fields above the face-center of the
cube are equivalent to the electric field of an infinitely charged plane.
Therefore, when $h\ll L$ and the NV-center is near the center of the cube face, the
amplification is independent of $h$. On the other hand, the stray field close to
the cube edge in comparison to $L$ is equivalent to the electric field of a set
of infinite line charges. Therefore, there is a logarithmic dependence of the
stray field on the distance to the edge, $d$, of the cube in units of
$L$ so that the amplification is weakly dependent on $d$.

Typical values of the stray fields at the position of the qubit and NV-center in the
limit $d,h\ll L$ for YIG are on the order of a few hundreds of Gauss. While the
presence of a magnetic field perpendicular to the NV-center axis can significantly
limit the read-out fidelity of the NV-center, it was found that the fields up to
$10$~mT can be tolerated.~\cite{tetienne_magnetic_2012} Therefore, we conclude
that one can either place the NV-center at distances where the stray field does
not exceed $10$~mT or care must be taken that the stray field is mostly aligned
along the NV-center axis.

\subsection{Estimates}
\label{sec:est}
In this section we give estimates for the amplification for two cases: detecting
a nuclear spin and an electron spin. For all the estimates provided herein, we
assume room temperature and that the FM material is YIG, so that
$\alpha\sim10^{-5}$, $\mu_0M_F=0.185$~T, and $K/M_F=60$~mT.~\cite{schryer_the_1974}
For simplicity but without loss of generality, we assume that the FM has
the shape of a cube for the estimates given below. For a cube
and in the macrospin approximation there is no contribution from shape
anisotropy. Because in typical experiments~\cite{grinolds_subnanometre_2014}
$h\sim5$~nm and $d=20-30$~nm, we take these values for our estimates below.

For detecting a single nuclear spin, we need to tune the FMR to the MHz range
and thus, according to Eq.~(\ref{eq:volumineq}), the optimal FM volume $V$
corresponds to a cube side $L=400-500$~nm. In this regime $d,h\ll L$, and thus
the amplified sensitivity only weakly depends on the particular choice of $d$
and $h$. Taking $T_2\sim200$~$\mu$s, which is the best case scenario for the
unamplified sensitivity and subsequently worst case for amplification, we obtain
$\nu\sim1.2\times 10^{3}$.  Since state-of-the-art NV-magnetometers can
resolve clusters of about 300 nuclear spins at a distance of
$\sim10$~nm,~\cite{Balasubramanian_Nanoscale_2008,Maze_Nanoscale_2008,Mamin_Nanoscale_2013,Staudacher01022013}
the amplification by three orders of magnitude obtained herein suffices for
resolving a single nuclear spin at a distance of $30$~nm from the surface. 

We note that the volume considered in our estimate corresponds to
$T_2^\prime\sim10$~ns [see Eq.~(\ref{eq:deltaphi})], which is also the value of
the optimal interrogation time, and therefore even when the ``bare'' $T_2\sim$
10~ns ({\it i.e.},  the $T_2$ time in the absence of the FM), the decoherence
time of the NV-center is limited by the FM noise. Thus, because the signal
amplification in our scheme far exceeds the effect of the additional decoherence
it induces, even shallow NV-centers~\cite{myers_probing_2014,ohno_engineering_2012}
with relatively short decoherence time can be used and significantly outperform
unamplified long-lived NV-centers. 

To amplify the signal from a single electron spin, there is no need to tune down the
FMR frequency since the electron spin can be driven at GHz frequencies. Thus, one
can use the stable state of the FM  and therefore there is no
restriction on the minimum FM volume. With the same distances $d$ and $h$ as in
the nuclear spin case and  for $L=20-40$~nm, corresponding to the volume of
maximum amplification, we obtain $\nu\sim 0.5\times 10^{4}$. In order to obtain
this estimate, we used the fact that the relaxation time of the electron spin is
typically shorter than the NV-center $T_2$ time and that the NV-center interrogation time
cannot be longer than the relaxation
time of the electron spin.~\cite{grinolds_subnanometre_2014} Thus, in Eq.~(\ref{eq:etaV}) we use
a typical electron spin relaxation time $T_1^d\sim150$~ns in place of
$T_2$.~\cite{grinolds_subnanometre_2014} If we choose $h=20$~nm and $d=1$~nm, we
obtain an amplification as big as $\nu\sim2\times10^4$ for $L=15-20$~nm.

\section{Conclusions}
We have proposed and analyzed, both analytically and numerically, a modification
of a standard NV-magnetometry setup that yields a significant improvement of
NV-magnetometer sensitivity by up to four orders of magnitude. Our scheme is
based on a ferromagnetic particle, placed in close proximity to a sensing
NV-center spin. The qubit spin to be detected is then used to resonantly drive
the large macrospin of the FM  giving rise to a strong, amplified stray field
acting on the NV-magnetometer. Compared to the existing schemes that use the
quantum nature of an intermediate spin for improving
sensitivity,~\cite{schaffry_proposed_2011} we stress that our scheme is fully
classical and thus should be easily realizable at room temperature---all the
ingredients of our scheme are already demonstrated in separate
experiments.~\cite{fuchs_gigahertz_2009,grinolds_subnanometre_2014,nolte_tracking_2014,wolfe_off_2014} 

An alternative setup to achieve resonance between the qubit and FM is to place
the NV-center and the FM on a cantilever~\cite{PhysRevLett.113.020503} with
resonance frequency in the GHz range. By driving the cantilever, we alleviate
the necessity of driving the qubit at FMR frequency as the qubit field is
modulated by the oscillations of the cantilever. Since the dipolar field of the
qubit decays rapidly with distance, the modulation of the qubit field achieved
in this scheme is almost as big as when the qubit is driven via a microwave
field (the previously described scheme for which the amplification estimates are
given). Therefore, we conclude that the estimates for the sensitivity
amplification given in Sec.~\ref{sec:est} still hold in that case.

The magnetometric scheme including a ferromagnetic particle proposed here is a
step forward to a more accurate magnetic field measurement. In particular, it
enables the detection of a single nuclear spin at distances that are noninvasive
to the system under study. Therefore, the proposed room temperature
amplification method opens up new venues for future analyses of previously
inaccessible biological and chemical systems.

\section{Acknowledgments}
We would like to thank H.~Fanghor for sharing his expertise about micromagnetic
simulations. This work was supported by the SNF, NCCR QSIT, IARPA, DARPA QuASAR
programs and the MURI QuISM. F.~L.~P. is grateful for support from the Alexander
von Humboldt foundation.
\appendix
\section{Cramer-Rao Bound}\label{sec:CR_App}
For the sake of completeness, we review here the proof of the Cramer-Rao bound,
\begin{align}
  \langle(\delta\hat\mu_s) ^2\rangle\ge 1/F(\mu_s),
  \label{eq:CR_App}
\end{align}
that we use to derive the sensitivity expression (\ref{eq:Sensitivity}) in the
main text. 

The Fisher information of the parameter estimator $\hat{\mu}_{s}$ is
given by
\begin{equation}
F(\mu_s)=-\sum_{n=\pm1}
  p(n\vert\mu_s)\frac{\partial^2\ln(p(n\vert\mu_s))}{\partial
  \mu_s^2}\,.
\end{equation}
The Cramer-Rao bound follows from the trivial identity
\begin{align}
  0&=\sum_{n_1}\dots\sum_{n_{\cal N}} p(n_1\vert\mu_s)\dots p(n_{\cal N}\vert\mu_s)\Delta\hat\mu_s,
  \label{eq:CRp1_app}
\end{align}
where $\Delta\hat\mu_s=\hat\mu_s(n_1,\dots,n_{\cal N})-\langle\hat\mu_s\rangle$.
Taking the derivative of this identity with respect to $\mu_s$ and using the
fact that the estimator $\hat\mu_s$ does not depend explicitly on $\mu_s$, we obtain

\begin{align}
  \sum_{n_1}&\dots\sum_{n_{\cal N}} p(n_1\vert\mu_s)\dots p(n_{\cal N}\vert\mu_s)\notag\\
  &\times\left( \sum_{k=1}^{\cal N}
  \frac{\partial\ln{p(n_k\vert\mu_s)}}{\partial\mu_s}
  \right)\Delta\hat\mu_s=\frac{d\langle\hat\mu_s\rangle}{d\mu_s}.
  \label{eq:CRp2_app}
\end{align}
Furthermore, for the unbiased estimator, $\langle\hat\mu_s\rangle=\mu_s$ and
thus the right-hand side of Eq.~(\ref{eq:CRp2_app}) is equal to $1$. Finally,
applying the Schwarz inequality,
$\mathrm{cov}(X,Y)^2\ge\mathrm{var}(X)\mathrm{var}(Y)$, to the above equation
yields the Cramer-Rao bound, Eq.~(\ref{eq:CR_App}).

\section{Sensitivity of an NV-center}\label{sec:Sensitivity_App}
The variance of any estimator $\hat{\mu}_{s}$ of the unknown parameter $\mu_s$
satisfies the Cramer-Rao inequality Eq.~(\ref{eq:CR_App}).
%
Using the probability distribution from
Eq.~(\ref{eq:pnphi}) we obtain

\begin{equation}
  \label{eq:Fisher_App}
  F(\mu_s)
  =\frac{(\partial\varphi_\mathrm{NV}(t_i)/\partial\mu_s)^2\sin^2(\varphi_\mathrm{NV}(t_i)+\theta)}{e^{2\langle(\delta\varphi_\mathrm{NV}(t_i))^2\rangle}-\cos^2(\varphi_\mathrm{NV}(t_i)+\theta)}\,.
\end{equation}
Thus, a bigger Fisher information $F(\mu_s)$ leads to a more accurate value of the
estimator $\hat{\mu}_{s}$.

For DC magnetometry we typically have
$\varphi_\mathrm{NV}(t_i)=\gamma_\mathrm{NV}(B_0+B)t_i\equiv t_i/T^\prime$ and
$\langle(\delta\varphi_\mathrm{NV}(t_i))^2\rangle=t_i/T_2^*$, where
$\gamma_\mathrm{NV}B_0=2.87$GHz is the zero-field splitting of the NV-center and
$T_2^*$ is typically on the order of a few microseconds.  The field we want to
measure is $B=\mu_0\mu_s/(4\pi d^3)$, with $d$ the distance between the qubit
and the NV-center.  Consider the scenario $T^\prime\ll T_2$ which is valid for DC
magnetometry. In such a situation, one may choose an interrogation time
maximizing $\sin^2[\varphi_\mathrm{NV}(t_i)+\theta]$ and minimizing
$\cos^2[\varphi_\mathrm{NV}(t_i)+\theta]$ in Eq.~(\ref{eq:Fisher_App}). This is
independent of the angle $\theta$ which we set to zero without loss of
generality. Thus instead of maximizing the Fisher information from
Eq.~(\ref{eq:Fisher_App}), we need only to maximize its envelope function, 

\begin{align}
  \label{eq:Fisherenvelope_App}
  \tilde
  F(\mu_s)&=(\partial\varphi_\mathrm{NV}(t_i)/\partial\mu_s)^2e^{-2\langle(\delta\varphi_\mathrm{NV}(t_i))^2\rangle}\,.
\end{align}
Repeating the measurement ${\cal N}=T/(t_p+t_i)$ times ($t_p$ is the
initialization time) reduces the variance by a factor $1/{\cal N}$. The Cramer-Rao
bound (\ref{eq:CR_App}) then leads to $\langle(\delta\hat\mu_s) ^2\rangle\ge
\frac1{ {\cal N}F(\mu_s)} \label{eq:CRN}$. The minimal value of the magnetic
moment $\tilde\mu_s(t_i,T)$ (that can be resolved within measurement time $T$ and
interrogation time $t_i$) is determined by the one for which the mean value is
equal to its standard deviation, 
\begin{equation}
\tilde\mu_s(t_i,T)=\frac1{\sqrt{ {\cal
N}F(\tilde\mu_s)}}\,.
\label{eq:tildemustT_app}
\end{equation}
Assuming $t_i\gg t_{p}$, we finally obtain the sensitivity
\begin{equation}\label{eq:Sensitivity_App}
  S=\frac1{R\sqrt{\eta}}\min_{t_i}\left[\frac{
  e^{\langle(\delta\varphi_\mathrm{NV}(t_i))^2\rangle}\sqrt{t_i}}{\vert
    \partial\varphi_\mathrm{NV}(t_i)/\partial\mu_s\vert}\right].
\end{equation}
As mentioned in the main text, $R$ is the measurement contrast and $\eta$ is the
detection efficiency;~\cite{PhysRevB.80.115202} these quantities take into
account that the measurement has to be performed many times in order to detect a
photon.

The situation for AC magnetometry is different; here
$\varphi_\mathrm{NV}(t_i)=\lambda\gamma_{NV}Bt_i\equiv t_i/T^\prime$ while we
still have $\langle(\delta\varphi_\mathrm{NV}(t_i))^2\rangle=t_i/T_2$. (Note
that the constant of proportionality $\lambda$ depends on the specific pulse
sequence applied.) In typical situations, the AC magnetic field is small and
$T^\prime\gg T_2$. In such scenario, and for vanishing angle $\theta$, the
accumulated phase will never reach a value of $\pi/2$ and one needs to maximize
the Fisher information (\ref{eq:Fisher_App}), not only its envelope. In this
limit, we obtain a very different expression for the sensitivity, and, in
particular, the expression is in units of ``magnetic
moment''$/\mathrm{Hz}^\frac14$ (see Appendix~\ref{sec:Sensitivity2_App}).
Fortunately, such result can be improved: one may take a nonzero value of the
angle $\theta$ such that the expression
$\varphi_{\mathrm{NV}}(t_i)+\theta=\pi/2$ within the interrogation time. In this
case, the amplified sensitivity takes the form (\ref{eq:Sensitivity_App}),
improving the sensitivity. 

\section{AC sensitivity for $\theta=0$}\label{sec:Sensitivity2_App}
The goal of this appendix is to derive the expression for the AC sensitivity
when the angle $\theta$ between the two $\pi/2$-pulses, applied at the beginning
and at the end of the sequence, is zero. This calculation is presented for the
sake of completeness, however, this is not the expression we use to derive our
amplified sensitivity. As mentioned in the main text, for AC magnetometry we
have

\begin{align}
  \varphi_\mathrm{NV}(t_i)&\propto\gamma_\mathrm{NV}Bt_i\equiv t_i/T^\prime,\\
  \langle(\delta\varphi_\mathrm{NV}(t_i))^2\rangle&=t_i/T_2,
  \label{eq:ACmagnetometry_app}
\end{align}
where the constant of proportionality depends on the specific pulse sequence
applied. It is important to note that here only the AC field component that
matches the frequency of the pulse sequence contributes ({\it i.e.}, there is no
contribution from $B_0$). When the magnitude of the AC driving field is small, we
are in the limit of $T^\prime\gg T_2$. As noted above, in this regime the
accumulated phase $\varphi_{\mathrm{NV}}(t_i)$ will not reach the value of
$\pi/2$ and thus the full Fisher information needs to be maximized. Namely, we
maximize
\begin{equation}
 F(\mu_{s})=\frac{(\partial\varphi_\mathrm{NV}(t_i)/\partial\mu_s)^2\sin^2(\varphi_\mathrm{NV}(t_i)+\theta)}{e^{2\langle(\delta\varphi_\mathrm{NV}(t_i))^2\rangle}-\cos^2(\varphi_\mathrm{NV}(t_i)+\theta)}\,.
\end{equation}
Therefore,
\begin{align}
  \sqrt{\langle(\delta\hat\mu_s)
  ^2\rangle}\ge\frac{\sqrt{
  e^{2\langle(\delta\varphi_\mathrm{NV}(t_i))^2\rangle}-\cos^2(\varphi_\mathrm{NV}(t_i))}\sqrt{t_p+t_i}}{\sqrt{T}\vert\sin(\varphi_\mathrm{NV}(t_i))\partial\varphi_\mathrm{NV}(t_i)/\partial\mu_s\vert}.
  \label{eq:Smu_app}
\end{align}
In contrast to the result obtained in Ref.~\onlinecite{schaffry_proposed_2011}
[which is similar to the one obtained in Eq.~(\ref{eq:Fisherenvelope_App})], the
Fisher information in Eq.~(\ref{eq:Smu_app}) depends on the estimation parameter
$\mu_s$. The minimal value of the magnetic moment $\tilde\mu_s(t_i,T)$ (that can
be resolved within measurement time $T$ and interrogation time $t_i$) is again
determined as the one for which the mean value is equal to its standard
deviation, Eq.~(\ref{eq:tildemustT_app}).
%
Therefore, using the fact that $\varphi_\mathrm{NV}(t_i)$ depends linearly on
$\mu_s$ and that for typical interrogation time  $\varphi_\mathrm{NV}(t_i)\ll1$,
we find an approximate solution to Eq.~(\ref{eq:tildemustT_app}). Minimizing over the
interrogation time, we obtain
\begin{align}
  \tilde\mu_s(T)&\equiv\min_{t_i}\left[\tilde\mu_s(t_i,T)\right]\\
  &=\min_{t_i}\left[\frac{(
  e^{2\langle(\delta\varphi_\mathrm{NV}(t_i))^2\rangle}-1)^{\frac14}}{\vert
  \partial\varphi_\mathrm{NV}(t_i)/\partial\mu_s\vert
  }\left( \frac{t_i}{T} \right)^\frac14\right].\notag
  \label{eq:tildemusT_app}
\end{align}
Now, if we remove the dependence on the total measurement time from the above
expression, we obtain the quantity that describes the magnetic moment sensitivity in
units of ``magnetic moment''$/\mathrm{Hz}^\frac14$, {\it i.e.},

\begin{equation}
  S=\frac1{R\sqrt{\eta}}\min_{t_i}\left[\frac{(
  e^{2\langle(\delta\varphi_\mathrm{NV}(t_i))^2\rangle}-1)^{\frac14}t_i^{\frac14}}{\vert
    \partial\varphi_\mathrm{NV}(t_i)/\partial\mu_s\vert}\right]\,.
  \label{eq:S_app}
\end{equation}

\section{Calculation of  $\varphi_{NV}(t_i)$ and $\beta(t_i,\tau)$}\label{sec:cal_App}
The goal of this appendix is to give a detailed derivation of
Eqs.~(\ref{eq:phiNVts}) and (\ref{eq:deltaphi}) that are central to our work.
The former describes the phase accumulated by the NV-magnetometer, while the
latter the variance of this accumulated phase.

Within linear response, the accumulated phase is
\begin{align}
  \varphi_\mathrm{NV}(t_i)=&\frac{\mu_s\gamma\gamma_\mathrm{NV}}{M_FV}\mathrm{Re}\left[iX_\xi(t_i,t^\prime)\left(\bm
  B_{F,s}^+\cdot\bm n_s\right)\left(\bm B_{F,\mathrm{NV}}^-\cdot \bm
  n_\mathrm{NV}\right)\right]\,,
  \label{eq:phi_acc_X}
\end{align}
where $\xi$ ($\vert\xi\vert\le\tau$) is the time offset between the CPMG pulse
sequence applied to the qubit and the NV-center. We have introduced the following
notation,

\begin{align}
  X_\xi(t_i,t^\prime)=&\int_0^{t^\prime}ds
  e^{-\Omega
  s}\int_0^{t_i}dt^{\prime\prime}f_\tau(t^{\prime\prime})f_\tau(t^{\prime\prime}-s-\xi)+\notag\\
  \label{eq:eqXttp}
  &+e^{-\Omega t^\prime}\int_0^{t_i} dse^{-\Omega s}p_\tau(s,t; \xi)\\
  \equiv&\tilde X_\xi(t_i,t^\prime)+e^{-\Omega t^\prime}Y_\xi(t_i),
\end{align}
with $\Omega=i\omega_F+\Gamma$ and $p_\tau(s,t_i;\xi)=\int_0^{t_i-s}dt^{\prime\prime}
f_\tau(t^{\prime\prime}-\xi)f_\tau(t^{\prime\prime}+s)$. After performing the
integral in Eq.~(\ref{eq:eqXttp}) and using $t_i=N\tau$,
$t^\prime=N^\prime\tau$, we obtain 

\begin{align}
  \label{eq:X}
  \tilde X_\xi(N\tau,N^\prime\tau)&=e^{-\Omega\xi}\left[
  \tilde{X}_{\xi=0}(N\tau,N^\prime\tau)\right.\nonumber\\
  &\left.-(1-e^{-\Omega
  t^\prime})\frac{\Omega\tau-2+e^{-\Omega\xi}(2+2\Omega\xi-\Omega\tau)}{\Omega^2}N
  \right],\\
  \tilde{X}_{\xi=0}(N\tau,N^\prime\tau)&=\frac{(1-e^{-N^\prime{\Omega\tau}})N(2+e^{\Omega\tau}({\Omega\tau}-2)+{\Omega\tau})}{\Omega^2(1+e^{\Omega\tau})}\,,\nonumber\\ \\
  Y_{\xi=0}(N\tau)&=-\frac{4\sh^4({\Omega\tau}/4)}{\Omega^2\ch^2({\Omega\tau}/2)}+\frac{2+{\Omega\tau}+({\Omega\tau}-2)e^{\Omega\tau}}{\Omega^2(1+e^{\Omega\tau})}N\nonumber\\
  &+\frac{(e^{{\Omega\tau}/2}-1)^4}{\Omega^2(1+e^{\Omega\tau})^2}e^{-N{\Omega\tau}}\,.
\end{align}
Since we want the qubit to perturb the ferromagnet within a narrow frequency
window around the FMR ({\it i.e.},  narrower than the FMR linewidth $\Gamma$),
we require that $\Gamma t^\prime\gg1$. In this limit, the expression for
$X_\xi(t_i,t^\prime)$ is significantly simplified
\begin{align}
  X_\xi(N\tau,N^\prime\tau)&=e^{-\Omega\xi}\left[
  \tilde{X}_{\xi=0}(N\tau,N^\prime\tau)\right.\\
  &\left.-\frac{\Omega\tau-2+e^{-\Omega\xi}(2+2\Omega\xi-\Omega\tau)}{\Omega^2}N
  \right],\nonumber\\
  \tilde X_{\xi=0}(N\tau,N^\prime\tau)&\approx X_{\xi=0}(N\tau)\\
  &\equiv\frac{N(2+e^{\Omega\tau}({\Omega\tau}-2)+{\Omega\tau})}{\Omega^2(1+e^{\Omega\tau})}\notag.
  \label{eq:XN}
\end{align}

The expression for dephasing can be obtained from~\cite{cywinski_how_2008}
\begin{equation}
  \beta(t_i,\tau)=\gamma_\mathrm{NV}^2\int_0^{t_i}ds\langle
  B_\mathrm{NV}(s)B_\mathrm{NV}(0)\rangle p_\tau(s,t_i; \xi=0).
  \label{eq:dephasin}
\end{equation}
Furthermore,
\begin{equation}
\langle B_\mathrm{NV}(t)B_\mathrm{NV}(0)\rangle=\vert
\bm B_{F,\mathrm{NV}}^+\cdot \bm n_\mathrm{NV}\vert^2\mathrm{Re}[\langle
m^+(t)m^-(0)\rangle].
  \label{eq:mpmm}
\end{equation}
In the limit $M_FV(b_a \pm b)\gg k_B T$ ($b_a=K/M_F$ is the anisotropy field)
one obtains the following expression for the fluctuations of the ferromagnet,
\begin{equation}
  \langle m^+(t)m^-(0)\rangle=\frac{2k_B
  T}{M_FV(b_a\pm b)}e^{-i\omega_Ft-\Gamma\vert t\vert}.
  \label{eq:fluctuations}
\end{equation}
Combining equations (\ref{eq:dephasin}-\ref{eq:fluctuations}) we finally obtain
\begin{align}
  \beta(N\tau,\tau)=&\frac{2\gamma_\mathrm{NV}^2\vert \bm B_{F,\mathrm{NV}}^+\cdot \bm
  n_\mathrm{NV}\vert^2k_B
  T}{M_FV(b_a\pm b)}Y_{\xi=0}^{\prime}(N\tau).
  \label{eq:gamma}
\end{align}

\subsection{On-resonance case $\tau=(2k+1)\pi/\omega_F$}\label{sec:onresonance}
The Fourier transform of the CPMG sequence depicted in Fig.~\ref{fig:sequence} has peaks at frequencies
$(2k+1)\pi/\tau$. Thus we have a resonant behavior whenever this frequency matches
$\omega_F$. Assuming $\Gamma t^\prime\gg1$ and $\omega_F\gg(2k+1)\Gamma$ leads
to the following expression for $X_\xi(t_i)$, namely 

\begin{align}
  X_\xi(N)&=\left[-\frac{4N}{(2k+1)\pi\Gamma\omega_F}+i\frac{\left( (2k+1)^2\pi^2-8
  \right)N}{(2k+1)\pi\omega_F^2}\right]e^{-i\psi},
  \label{eq:Xt}
\end{align}
where $\psi=\omega_F\xi$. Assuming $\omega_F\gg(2k+1)^2\pi^2\Gamma$, we can
neglect the second term in the bracket in Eq.~(\ref{eq:Xt}) and obtain the
expression for the phase accumulated by the NV-center during the interrogation time
$t_i$,

\begin{align}
  \varphi_\mathrm{NV}(t_i)&=\frac{4\mu_s\gamma\gamma_\mathrm{NV}\mathrm{Im}\left[
  e^{-i\psi}(\bm B_{F,s}^+\cdot\bm n_s)(\bm B_{F,\mathrm{NV}}^-\cdot \bm
  n_\mathrm{NV})\right]}{\pi^2(2k+1)^2M_FV\Gamma}t_{i}\,.
  \label{eq:phiNVts_App}
\end{align}
Next, we choose the time offset $\xi$ ({\it i.e.}, $\psi$) such that
$\varphi_\mathrm{NV}(t_i)$ in the above equation is maximized
\begin{equation}
  \psi=\arg\left[(\bm B_{F,s}^+\cdot\bm
  n_s)(\bm B_{F,\mathrm{NV}}^-\cdot \bm n_\mathrm{NV})\right]
\end{equation}
we obtain
\begin{align}
  \varphi_\mathrm{NV}(t_i)&=\frac{4\mu_s\gamma\gamma_\mathrm{NV}
  \vert\bm B_{F,s}^+\cdot\bm n_s\vert\vert\bm B_{F,\mathrm{NV}}^-\cdot \bm
  n_\mathrm{NV}\vert}{\pi^2(2k+1)^2M_FV\Gamma}t_{i}\,.
  \label{eq:phiNVts_App}
\end{align}
Assuming that the optimal interrogation time satisfies $\Gamma t_i\ll1$, we
arrive at the following expression for $Y_{\xi=0}^\prime(t_i)$, namely
\begin{align}
  Y_{\xi=0}^\prime(N)&=\frac{2N}{\omega_F^3}\left[ \left( (2k+1)\pi-4(-1)^k
  \right)\Gamma+N\omega_F \right]\notag\\
  &\sim\frac{2N^2}{\omega_F^2}.
  \label{eq:YN}
\end{align}
The above expression yields the following variance of the phase accumulated by
the NV-center

\begin{align}
  \beta(t_i,\tau)&=\frac{4\gamma\gamma_\mathrm{NV}^2\vert
  \bm B_{F,\mathrm{NV}}^+\cdot \bm
  n_\mathrm{NV}\vert^2k_BT}{\pi^2(2k+1)^2M_FV\omega_F}t_i^{2}\,.
  \label{eq:deltaphi_App}
\end{align}

\subsection{Pulse sequence applied: a matter of
timescales}\label{sec:pulsesequence}\label{sec:D2}

\begin{figure}
  \centering\includegraphics[width=\columnwidth]{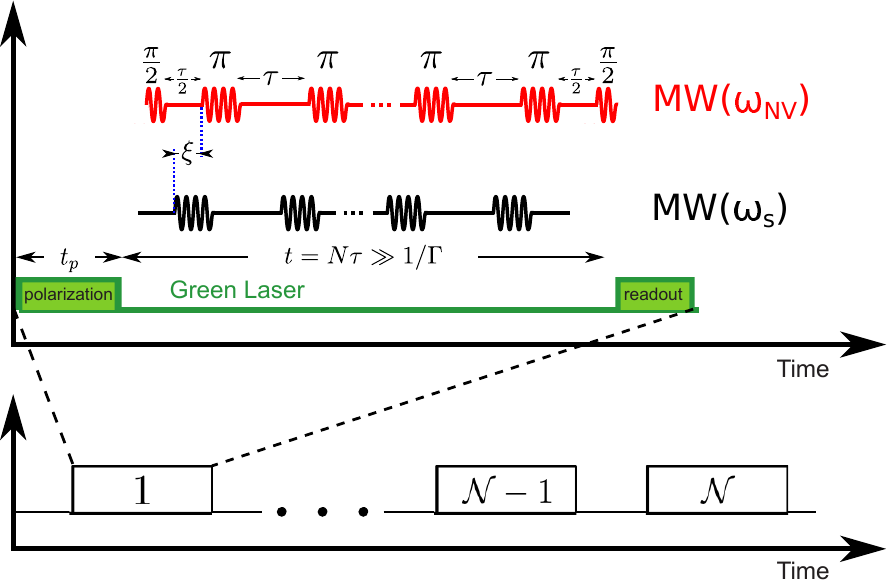}
  \caption{One choice for the pulse sequence applied to the qubit (black) and
  the NV-center spin (red). The pulse sequence $f_\tau(t)$ that consists of $2N$
  pulses is applied to both spins, with the time offset $\xi$, during the
  interrogation time $t$. The measurement is repeated $\cal N$ times until the
  desired precision is achieved, as illustrated on the bottom panel. We assume
  that the frequencies $\omega_s$, $\omega_\mathrm{NV}$, and $\omega_F$ are all
  different.}
  \label{fig:sequence_equal}
\end{figure}
Here we elucidate the importance of the different duration of the pulse
sequences applied to the qubit and the NV-center. We analyze what happens to the
amplification when both pulse sequences have the same duration, see
Fig.~\ref{fig:sequence_equal}. Performing the Fourier transform of
Eq.~(\ref{eq:dephasin}), one obtains
\begin{align}
  \beta(t_i,\tau)=&\gamma_\mathrm{NV}^2\vert
  \bm B_{F,\mathrm{NV}}^+\cdot \bm
  n_\mathrm{NV}\vert^2\times\notag\\
  &\times\int\frac{d\omega}{2\pi}
  \langle m_-(t)m_+(0)\rangle_\omega\frac{F(\omega
  t_i)}{\omega^2}\notag\\
  \approx& \gamma_\mathrm{NV}^2\vert \bm B_{F,\mathrm{NV}}^+\cdot \bm
  n_\mathrm{NV}\vert^2\langle m_-(t)m_+(0)\rangle_{\omega_F} \times\notag\\
  &\times\int\frac{d\omega}{2\pi}\frac{F(\omega t_i)}{\omega^2}\notag\\
  =&\gamma_\mathrm{NV}^2\vert \bm B_{F,\mathrm{NV}}^+\cdot \bm
  n_\mathrm{NV}\vert^2\langle m_-(t)m_+(0)\rangle_{\omega_F} t_i\notag\\
  \equiv& t_i/T_2^\prime\,.
  \label{eq:NSRest}
\end{align}
Here the subscript $\omega_{F}$ refers to the Fourier transform evaluated at
frequency $\omega_{F}$. We assume that the filter function $F(\omega t_{i})$ is
centered around the frequency $\omega_F$ (on-resonance case) and that it is much
narrower than the FMR linewidth, {\it i.e.}, $\Gamma t_i\gg1$---this is exactly
the opposite limit from the one assumed to arrive at Eq.~(\ref{eq:deltaphi}).
The accumulated phase is given in Eq.~(\ref{eq:phiNVts}), and thus the magnetic
moment sensitivity in this case reads

\begin{align}
  S_A=\frac{\pi^2}{4i\, \vert\bm B_{F,s}^+\cdot\bm n_s\vert}\frac{\sqrt{\langle
  m_-(t)m_+(0)\rangle_{\omega_F}}}{\chi_\perp(\omega_F)},
  \label{eq:SAnoamp}
\end{align}
where $\chi_\perp(\omega)=\gamma/\left[ M_FV(\omega_F-\omega+i\Gamma) \right]$.
The above expression yields almost no amplification, since we do not only excite
the FM resonantly but also the NV-center picks up the resonant noise. If we
rewrite Eq.~(\ref{eq:deltaphi}) in the form $\beta(t_i,\tau)\equiv
(t_i/T_2^{\prime\prime})^2$, we can understand that the decoherence times in the
two considered limits differ from each other by many orders of magnitude, namely

\begin{align}
  \frac{ T_2^{\prime\prime}
  }{T_2^\prime}=\frac{\Gamma}{\pi\gamma_\mathrm{NV}\vert
  \bm B_{F,\mathrm{NV}}^+\cdot \bm n_\mathrm{NV}\vert}\sqrt{\frac{M_FV\omega_F}{\gamma
  k_BT}}\gg1.
  \label{eq:T2pT2pp}
\end{align}

\begin{figure}[t]
  \centering\includegraphics[width=\columnwidth]{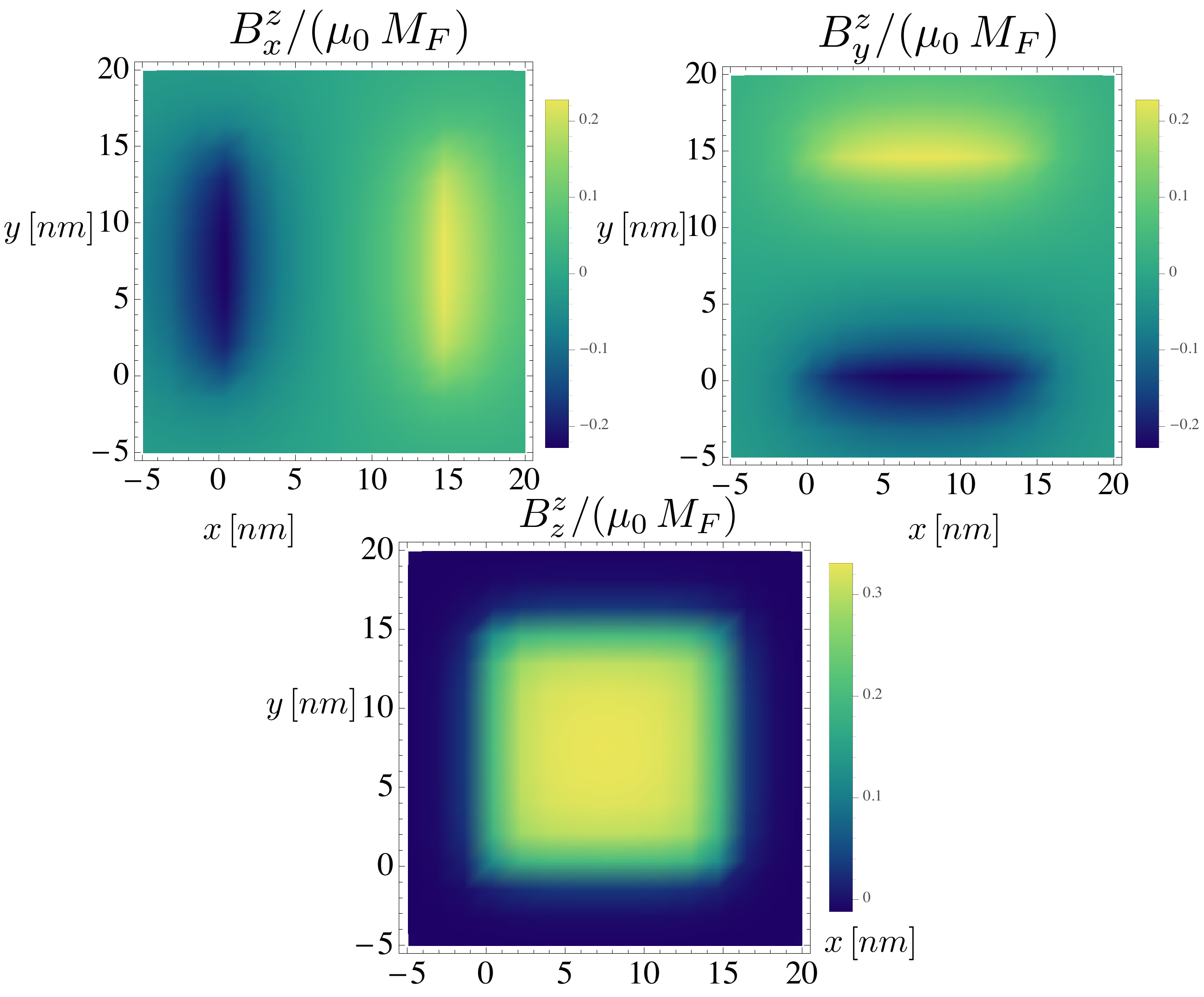}
  \caption{Plot of the stray field components normalized to the magnetic
  saturation: $B_{x}^{z}(x,y,z=16\,\textrm{nm})/(\mu_0 M_{F})$,
  $B_{y}^{z}(x,y,z=16\,\textrm{nm})/(\mu_0 M_{F})$, and
  $B_{z}^{z}(x,y,z=16\,\textrm{nm})/(\mu_0 M_{F})$ for a cube of side length
  $L=15\,\textrm{nm}$. Because the origin of our coordinate system is at the
  center of the cube, this is a plot of the stray field at a distance of
  $1\,\textrm{nm}$ from the upper surface of the cube.}
  \label{fig:Stray}
\end{figure}

\section{Stray field from a uniformly magnetized cuboid}\label{app:Stray}

In this section we review the analytical formulas giving the stray field of a
uniformly magnetized cuboid of side lengths $L_{x}$, $L_{y}$, and $L_{z}$, see
Fig.~\ref{fig:geometry}. As mentioned in the main text, the magnetic field
$\mathbf{B}({\bf r})$ at a point ${\bf r}=(x,y,z)$ outside of the cuboid can be
calculated from the expression for the electric field originating from charges
uniformly distributed on the surfaces of the cuboid perpendicular to the
magnetization,~\cite{Norpoth,EngelHerbet} see Fig.~\ref{fig:geometry}. For the
sake of simplicity, we assume that the magnetization direction points either
along $x$, $y$, or $z$. The expression for the stray field is then
\begin{equation}\label{eq:stray}
\mathbf{B}^{\delta}({\bf
r})=\frac{\mu_{0}\,M_F}{4\pi}\int_{0}^{L_{\alpha}}d\alpha\int_{0}^{L_{\beta}}d\beta\left\{\frac{{\bf
r}-{\bf r}^{\alpha\beta}_{\delta}}{\vert {\bf r}-{\bf
r}^{\alpha\beta}_{\delta}\vert^{3}}-\frac{{\bf r}-{\bf
r}^{\alpha\beta}_{\bar{\delta}}}{\vert {\bf r}-{\bf
r}^{\alpha\beta}_{\bar{\delta}}\vert^{3}}\right\}
\end{equation}
for $\delta=x,y,z$. Here, $\alpha$ and $\beta$ are the
directions perpendicular to $\delta$, {\it i.e.}, ${\bf
r}^{\alpha\beta}_{x}=(L_{x},\alpha,\beta)$, ${\bf
r}^{\alpha\beta}_{y}=(\alpha,L_{y},\beta)$, and  ${\bf
r}^{\alpha\beta}_{z}=(\alpha,\beta,L_{z})$, and 
$r_{\bar{\delta}}^{\alpha\beta}=r_{\delta}^{\alpha\beta}\mid_{L_{\delta}=0}$.

The integrals in Eq.~(\ref{eq:stray}) can be evaluated
analytically.~\cite{Norpoth,EngelHerbet} When the cuboid is magnetized along
$z$, one obtains 
\begin{eqnarray}
&&B^{x}_{x}(x,y,z)=\frac{\mu_{0}\,M_F}{4\pi}\left\{f(x,y,z)-f(x,y-L_{y},z)\right.\nonumber\\
&&\hspace{3cm}\left.-f(x-L_{x},y,z)+f(x-L_{x},y-L_{y},z)\right\}\nonumber\\
\\
&&B^{x}_{y}(x,y,z)=\frac{\mu_{0}\,M_F}{4\pi}\left\{f(y,x,z)-f(y-L_{y},x,z)\right.\nonumber\\
&&\hspace{3cm}\left.-f(y,x-L_{x},z)+f(y-L_{y},x-L_{x},z)\right\}\nonumber\\
\\
&&B^{x}_{z}(x,y,z)=\frac{\mu_{0}\,M_F}{4\pi}\left\{g(x,y,L_{z},z)-g(x,y-L_{y},L_{z},z)\right.\nonumber\\
&&\hspace{1.5cm}-g(x-L_{x},y,L_{z},z)+g(x-L_{x},y-L_{y},L_{z},z)\nonumber\\
&&\hspace{1.5cm}-g(x,y,0,z)+g(x,y-L_{y},0,z)\nonumber\\
&&\hspace{1.5cm}\left.+g(x-L_{x},y,0,z)-g(x-L_{x},y-L_{y},0,z)\right\}\,.
\end{eqnarray}
Here
\begin{eqnarray}
f(a,b,z)&=&\log\left(\frac{\sqrt{a^2+(z-L_{z})^{2}}(b+\sqrt{a^2+b^2+z^2})}{\sqrt{a^2+z^2}(b+\sqrt{a^2+b^2+(z-L_{z})^{2}})}\right)\,,\nonumber\\\\
g(a,b,c,z)&=&\arctan\left(\frac{a\,b}{(z-c)\sqrt{a^{2}+b^2+(z-c)^{2}}}\right)\,.
\end{eqnarray}
The analytical expressions for $\mathbf{B}^{y,z}({\bf r})$ are found similarly. 

For the sake of illustration, we plot in Fig.~\ref{fig:Stray} the three
components of $\mathbf{B}^{z}$ as function of $x$ and $y$ for a cube of size
$L=L_x=L_y=L_z=15\,\textrm{nm}$ at a distance of $1\,\textrm{nm}$ above the
upper face.

\bibliography{amplifier}
\end{document}